\documentclass[aps,prl,superscriptaddress]{revtex4-2}

\usepackage{amsmath}
\usepackage{graphicx}
\usepackage{MnSymbol}

\begin{document}

\title{On the stability of spinning asteroids}

\author{B.N.J. Persson}
\affiliation{PGI-1, FZ J\"ulich, Germany, EU}
\affiliation{www.MultiscaleConsulting.com}
\author{J. Biele}
\affiliation{German Aerospace Center, 51147 K\"oln, Germany, EU}

\begin{abstract}
Most asteroids with a diameter larger than $\sim 300 \ {\rm m}$ are
rubble piles i.e. consisting of more than one solid object. All 
asteroids are rotating but almost all asteroids larger than $\sim 300 \ {\rm m}$
rotate with a period longer than $2.3 \ {\rm hours}$, which is the critical period
where the centrifugal force equals the gravitational force. This indicates 
that there are nearly no adhesive interaction forces between the asteroid fragments.
We show that this is due to the surface roughness of the asteroid particles which reduces
the van der Waals interaction between the particles by a factor of $100$ for micrometer
sized particles and even more for larger particles. We show that  surface roughness
results in an interaction force which is independent of the size of the particles, in contrast to the
linear size dependency expected for particles with smooth surfaces. Thus, two stone fragments of 
size $100 \ {\rm nm}$ attract each other with the same non-gravitational 
force as two fragments of size $10 \ {\rm m}$.
\end{abstract}

\maketitle

\setcounter{page}{1}
\pagenumbering{arabic}




\begin{figure}[tbp]
\includegraphics[width=0.40\textwidth,angle=0]{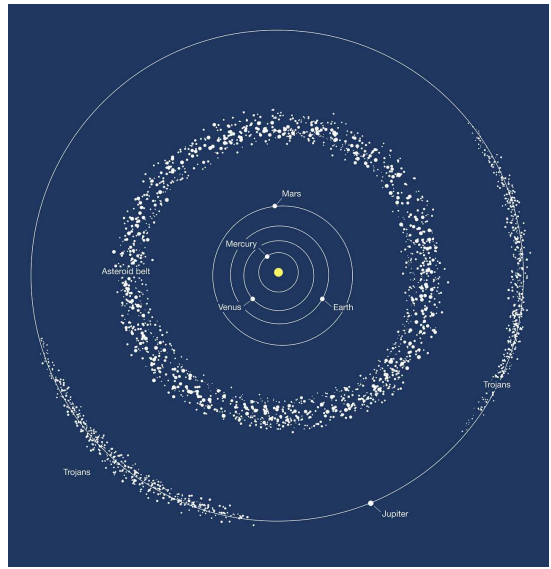}
\caption{
The asteroid belt between Mars and Jupiter was formed $\sim 4.6 \ {\rm billion}$
years ago, at the same time as the solar system was formed. 
Image credit: ESA/Hubble, M. Kornmesser.
}
\label{Ast1a.eps}
\end{figure}

\begin{figure}[tbp]
\includegraphics[width=0.65\textwidth,angle=0]{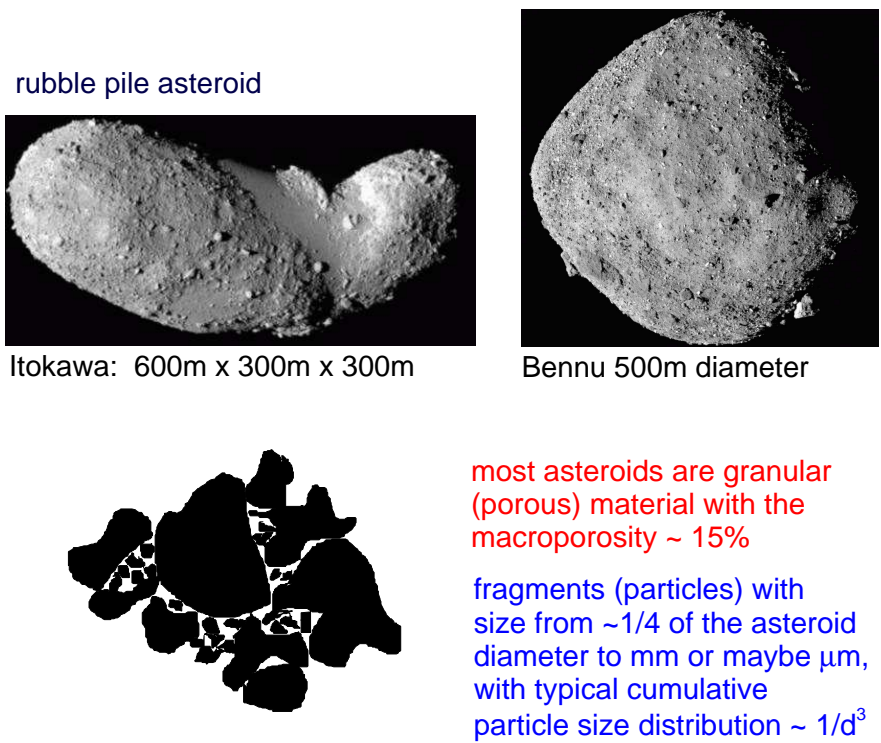}
\caption{
Most asteroids with a diameter $> 300 \ {\rm m}$ consist of many fragments
kept together mainly by gravity. They are denoted rubble pile asteroids and 
are non-spherical due to the weak gravitational field and by the fact that some fragments may
have a size similar to the size of the asteroid. 
Rubble pile asteroids consist of a broad range of fragment 
(or particle) sizes with a cumulative probability 
distribution which scales roughly as $1/d^3$ with the diameter $d$ of the particle \cite{metzger2020,michi21,burke21}.
The macro-porosity of rubble pile asteroids
is $\sim 15 \%$ (see \cite{Grott,Biele}). 
Asteroid Itokawa image credit: JAXA.
Asteroid Bennu image credits: NASA/Goddard/University of Arizona.
}
\label{Ast2.eps}
\end{figure}

\begin{figure}[tbp]
\includegraphics[width=0.65\textwidth,angle=0]{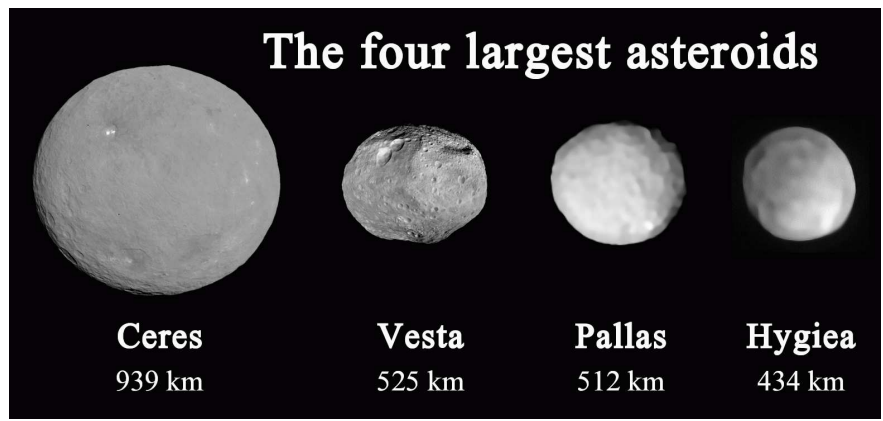}
\caption{
The 4 largest asteroids are nearly spherical due to gravity.
Ceres and Vesta images: NASA/JPL-Caltech/UCLA/MPS/DLR/IDA, Pallas and Hygiea images: ESO
}
\label{Ast1b.eps}
\end{figure}

\begin{figure}[tbp]
\includegraphics[width=0.2\textwidth,angle=0]{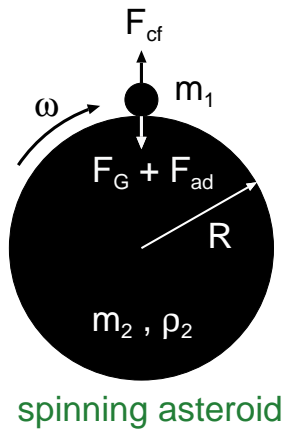}
\caption{
A particle located at the surface of an asteroid experiences a centrifugal force  and the gravity force,
and some attractive adhesion force to the surrounding contacting asteroid particles. 
}
\label{SphereCentri.a.eps}
\end{figure}

\begin{figure}[tbp]
\includegraphics[width=0.75\textwidth,angle=0]{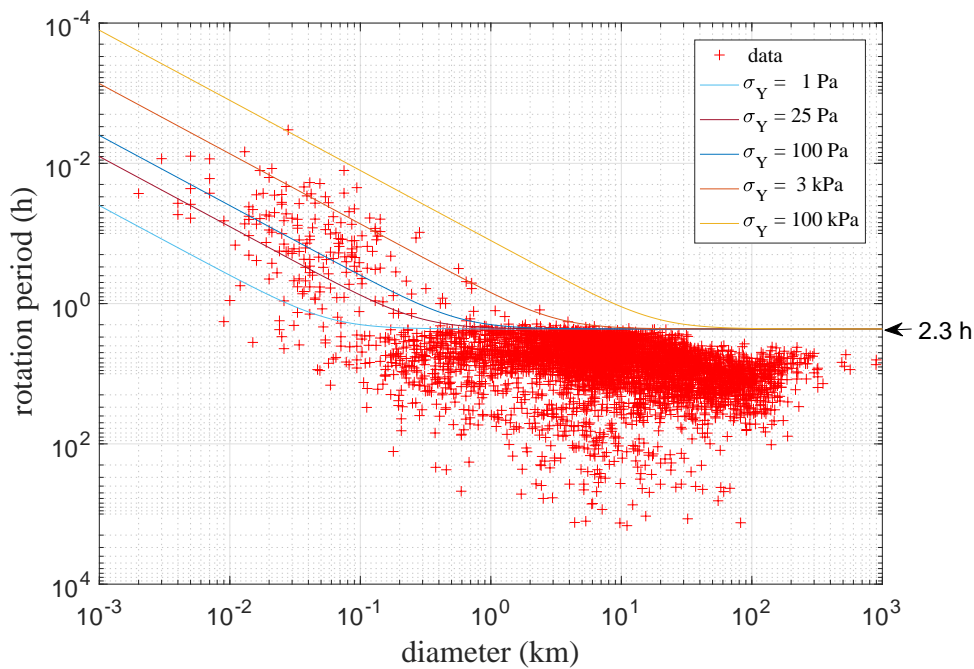}
\caption{
Nearly all asteroids
with the diameter $\sim 300 \ {\rm m}$ rotate with a period below $2.3 \ {\rm hours}$ but some smaller
asteroids rotate much faster. This is possible only if they consist of a single fragment,
or if they are rubble pile asteroids where the fragments are bound together with some adhesive bonds
in addition to the gravity attraction.
Asteroid data taken from \cite{lcdb}, all objects which have a PFlag or DiamFlag entry or a quality 
number $U < 2.5$ filtered out. Lines are calculated with Eqn. (24) of \cite{San} 
using bulk density $2100 kg/m^3$ and setting $\alpha=D/2$
}
\label{SphereCentri.b.eps}
\end{figure}

\begin{figure}[tbp]
\includegraphics[width=0.65\textwidth,angle=0]{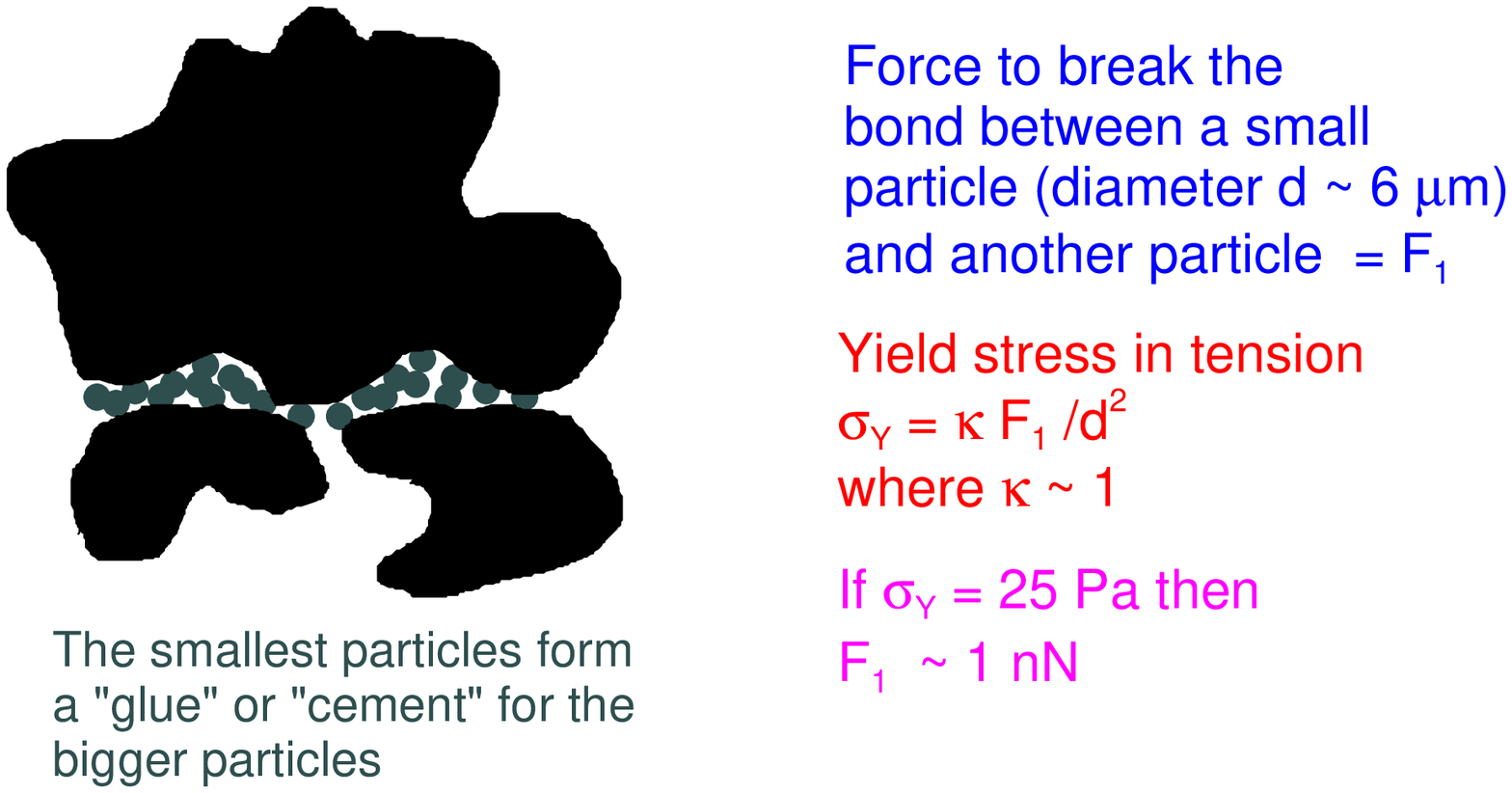}
\caption{
The big particles (fragments) in an asteroid are assumed to be kept together by a 
matrix of smaller particles. Analysis of experimental data gives an effective yield stress
of rubble pile asteroids of order (or less than) $\sigma_{\rm Y} \approx 25 \ {\rm Pa}$\cite{San}.
}
\label{GluedAstro.1a.eps}
\end{figure}

\vskip 0.3cm
{\bf 1 Introduction}

Asteroids are rocky, airless remnants left over from the early formation of our solar 
system about 4.6 billion years ago\cite{michel2015asteroids}. 
Most of this ancient space debris can be found orbiting our Sun between Mars and Jupiter 
within the main asteroid belt (see Fig. \ref{Ast1a.eps}). 
Early in the history of the solar system, the gravity of newly formed Jupiter brought an 
end to the formation of planetary bodies in this region and caused the small bodies to collide 
with one another, fragmenting them into the asteroids we observe today.

Asteroids range in size from Ceres--the largest at about 939 kilometers 
in diameter--to bodies that are less than 10 meters across. 
The most common asteroids consist of silicate rocks\cite{clay}.

Most asteroids are irregularly shaped (see Fig. \ref{Ast2.eps}), but a few of the biggest are nearly spherical 
due to the influence of gravity (see Fig. \ref{Ast1b.eps}). 
As they revolve around the Sun in (weakly) elliptical orbits, the asteroids also rotate. 
On the time-scale of million of years the rotation speed of asteroids change 
due to the momentum of photons (from the sun) absorbed, reflected 
and emitted (heat radiation) from the surface of asteroids\cite{rad}, and due to the impact of meteorites.
Most of the asteroids rotate with a period longer than $2.3 \ {\rm hours}$ which can be understood as a result of the
influence of the centrifugal force on a collection of solid fragments bound together mainly by the gravity force\cite{hestroffer2019small}.
However, many small asteroids rotate much faster. Many of these fast rotating asteroids may consist of single
solid blocks (monoliths) where the atoms are bound together with strong chemical bonds. However, even some 
of the small fast rotating asteroids are believed 
to be composite objects (gravitational aggregates)\cite{Sch}, which indicate
that in addition to gravity, some other weak force field must
act between the fragments or otherwise these asteroids would break-up 
due to the centrifugal force\cite{San}. One such force field is the van der Waals interaction, and we will
show below that taking this interaction into account gives results in agreement with experimental observations. 
However, it is necessary to include the surface roughness in the analysis, which was not
done in earlier studies.

Elastically stiff solid objects usually attract each other very weakly, 
and the force needed to separate two solid objects,
e.g. a glass bottle from a table, is usually so small that it cannot be detected without a very sensitive instrument.
The fundamental reason for this is surface roughness, which results in a very small contact area. In fact, in an ideal case,
for perfectly smooth surfaces, the van der Waals interaction is quite strong, e.g. it is possible to keep the weight of a car
with the van der Waals interaction acting over a surface area $\sim 1 \ {\rm cm}^2$ (see Ref. \cite{Rev,Ref5}). 
However, in practice this is never observed due to surface roughness and non-uniform bond breaking at the interface.

It is interesting to determine
the particle diameter $d$ where the mutual gravitational force $F_{\rm G} = G m_1 m_2/d^2$ would be equal to the non-gravitational force.
We will show below that the force between stone fragments are typically $\sim 1 \ {\rm \mu N}$ due to
capillary bridges and  $\sim 1 \ {\rm nN}$ due to the van der Waals interaction. 
For these two cases, using $m_1=m_2 = (\pi/6) \rho d^3 $ we get $d=6 \ {\rm cm}$ and $d=24 \ {\rm cm}$, 
respectively, where we have used the mass density $\rho = 2000 \ {\rm kg/m^3}$.

\begin{figure}[tbp]
\includegraphics[width=0.35\textwidth,angle=0]{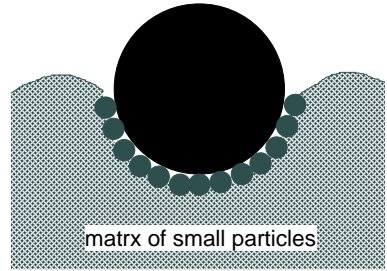}
\caption{
A big particle (fragment) bound to an asteroid via a matrix of smaller particles.
}
\label{Surrounded.eps}
\end{figure}

\vskip 0.3cm
{\bf 2 Strength of adhesion in composite asteroids}

Here we will present an approximate calculation of the angular rotation speed at break-up of rubble pile asteroids, the so called
spin barrier of gravitational aggregates. In reality, the ultimate structural failure mechanism after spin-up 
(which may include plastic-like deformation before rupture) is quite complex (global failure states, e.g. \cite{holsa07}). 
More quantitative calculations, including new data for meteorite fragment 
roughness power spectra, updated Hamaker constants and realistic strength models, will be 
presented in another paper.

Consider a particle with mass $m_1$, bulk density $\rho_1$  on the surface of a 
rotating body (asteroid) with the radius $R$, bulk density $\rho_2$ and mass $m_2$. 
If the particle is on the equator it will experience the centrifugal force
$$F_{\rm cf} = m_1 {v^2 \over R},$$
where $v=\omega R$ is the rotation speed. 
On the particle act the gravitational force $F_{\rm G}=G m_1 m_2/R^2$ 
and an adhesive force $F_{\rm ad}$ from the non-gravitational interaction with the asteroid
(see Fig. \ref{SphereCentri.a.eps}).
The condition for break-up is $F_{\rm G}+F_{\rm ad}=F_{\rm cf}$, or
using that $m_2 = (4\pi/3) R^3 \rho_2$:
$${4 \pi \over 3} G R m_1 \rho_2+ F_{\rm ad}=m_1 {v^2 \over R}$$
or
$$\omega = \left ({4 \pi \over 3} G \rho_2+ {F_{\rm ad}\over Rm_1}\right )^{1/2}\eqno(1)$$
If $F_{\rm ad}=0$ then the maximum angular rotation
velocity possible before breakup is determined by $F_{\rm G}=F_{\rm cf}$ giving
$\omega=\omega_0$ with
$$\omega_0 = \left ({4 \pi \over 3}G\rho_2 \right )^{1/2}\eqno(2)$$
This gives the rotation time period $T_0=2 \pi/ \omega_0$, or if we use the typical bulk density 
$\rho_2 = 2000 \ {\rm kg/m^3}$, the time period $T_0 \approx 2.3 \ {\rm hour}$ which is in good
agreement with observations for asteroids with the diameter $> 300 \ {\rm m}$ (see Fig. \ref{SphereCentri.b.eps}).
Using (1) and (2) we can write
$$\omega = \left (\omega_0^2+{F_{\rm ad}\over Rm_1}\right )^{1/2}\eqno(3)$$
Since the adhesion term in (3) depends on the radius of the asteroid as $1/R$ it follows that while
adhesion may be unimportant for large enough asteroids, it could be very important
for asteroids with small radius $R$.

We will show below that because of surface roughness the adhesion force $F_{\rm ad}$ does not depend
on the size of the particles, and for van der Waals (VDW) interaction it is typically 
$F_{\rm ad} \approx 10^{-9} \ {\rm N}$. Using this result and assuming first
that all the particles in an asteroid have the same size (say diameter $d_0$) we can estimate the maximum size
where adhesion matter. Thus if we assume
$${F_{\rm ad}\over Rm_1} \approx \omega_0^2 ,$$
the rotation frequency at break-up would be $\sqrt 2 \approx 1.4$ times bigger 
%
%
than in the absence of adhesion. 
Assuming $R\approx 100 \ {\rm m}$, $F_{\rm ad} = 10^{-9} \ {\rm N}$ and $\omega_0= 2.3 \ {\rm hour}$
this gives $m_1 \approx 10^{-5} \ {\rm kg}$ or, if the mass density is $2\times 10^3 \ {\rm kg/m^3}$,
the critical particle diameter $d_* \approx 1 \ {\rm mm}$.
If the particles would be bigger than this the asteroid would break-up (or emit particles) 
when rotating with $\omega \approx 1.4 \omega_0$.
However, asteroids consist of particles of different sizes and the smallest particles (if enough of them) could act as a glue
(cement) for the bigger particles\cite{San}. To study this consider the simplest case of a big particle (fragment) (diameter $d_1$) 
bound to other big particles (fragments) via a matrix of much smaller particles with diameter $d_0$ (see Fig. \ref{Surrounded.eps}). 
If half of the surface of the big particle is covered by 
the small particles then the number of bonds will be $\sim (d_1/d_0)^2$, and if all these bonds would break simultaneously the adhesion
force acting on the big particle would be $\sim (d_1/d_0)^2 F_{\rm ad}$. Replacing $F_{\rm ad}$ in (3) with this expression gives
$$\omega = \left (\omega_0^2+\left ({d_1\over d_0}\right )^2 {F_{\rm ad}\over Rm_1}\right )^{1/2}\eqno(4)$$
and the condition
$$\left ({d_1\over d_0}\right )^2 {F_{\rm ad}\over Rm_1} \approx \omega_0^2$$
gives $d_1 = d_*^3 / d_0^2$. 
As an example, if $d_* = 1 \ {\rm mm}$ and $d_0 = 0.1 \ {\rm mm}$, we get $d_1 \approx 10 \ {\rm cm}$.
However, rubble pile asteroids typically have fragments with linear size of order $\sim 10 \ {\rm m}$ and for such
asteroids not to break up until $\omega \approx 1.4 \omega_0$ one would need the ``cement'' particles to have the diameter
$d_0 \approx 10 \ {\rm \mu m}$.

Before break-up (global failure), a rubble pile asteroid may undergo plastic-like deformation where the asteroid fragments
change their relative positions. If we assume that the big fragments are bound together by the 
``cement'' of the smallest particles (with the diameter $d_0$) we expect from dimensional arguments a cohesive stress\cite{Rumpf,Pablo}
$$\sigma_{\rm Y} = \kappa {F\over d_0^2},\eqno(5)$$
where $\kappa$ is a number of order unity which depends on the porosity and the average number of 
particles touching one particle. Analysis of experimental data indicate that 
for rubble pile asteroids $\sigma_{\rm Y} \approx 25 \ {\rm Pa}$ (see dotted blue line in Fig. \ref{SphereCentri.b.eps})\cite{San}. 
Using (5) with $F\approx 10^{-9} \ {\rm N}$, 
$\sigma_{\rm Y} \approx 25 \ {\rm Pa}$ and $\kappa \approx 1$ we get $d_0 \approx 10 \ {\rm \mu m}$, 
which is the same as found above using another argument.

The particle size-distribution of rubble pile asteroids has been studied using optical observations
from satellite's for big particles, and from samples collected in the Hayabusa mission 
to the asteroid Itokawa\cite{metzger2020,michi21,burke21}. These studies indicate a
cumulative size distribution which scales as $\approx d^{-3}$ with the linear size $d$ of the particle
from $d \approx 100 \ {\rm m}$ down to a few ${\rm \mu m}$, which appears as the lower cut-off in the
probability distribution. However, recent observations 
indicate that the surface and shallow (dm to m) subsurface
 may be strongly deficient in fines with size $< 1 \ {\rm mm}$. However, these asteroids rotate with a period well above
$2.3 \ {\rm hours}$. The study in Ref. \cite{San} showed that there if the $d^{-3}$ number size distribution holds for
all particle sizes then there are enough small particles to
form a binding matrix (cement) surrounding the bigger stone fragments.

\begin{figure}[tbp]
\includegraphics[width=0.2\textwidth,angle=0]{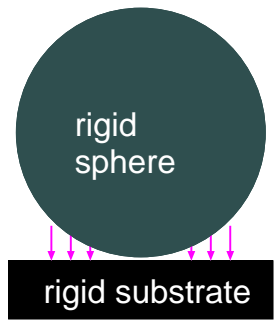}
\caption{
If the solids are assumed rigid the sphere 
will be in contact with the flat in a single point. The two objects attract each
other by a force field which is large only close to the contact point.
}
\label{GluedAstro.1b.eps}
\end{figure}

\vskip 0.3cm
{\bf 3 Adhesion of particles with smooth surfaces}

Consider the force to separate two solid particles in adhesive contact. 
The adhesion force $F_{\rm ad}$ could have several different origins, namely:

(a) Electrostatic effects.
Asteroids are exposed to ultraviolet radiation and ions (plasma) from the sun which can ionize or
charge particles in asteroids. However, this is likely to affect only a thin surface layer on the asteroids.

(b) Van der Waals interaction.  
Quantum (and thermal) fluctuations in the charge distribution in solids result in polarization effects
of the fluctuating dipole--induced dipole type. The interaction between the fluctuating dipoles 
is always attractive when the bodies are surrounded by vacuum, and act between all solid objects.
When two bodies are surrounded by a fluid the VDW interaction can be repulsive\cite{Is}.

(c) Capillary bridges. In the normal (humid) atmosphere capillary bridges usually gives
the dominant adhesion force, e.g., between sand particles.
However, asteroids are surrounded by (nearly) vacuum and it is unlikely that any mobile
molecules occur on the solid fragments which could rearrange (diffuse) and form capillary bridges.
To understand this note that according to Kramers theory\cite{Kramer} of thermally activated processes, 
the rate of jumping over a barrier of height $\epsilon$ is
$$w= {\omega_0\over 2 \pi} e^{-\epsilon /k_{\rm B }T}\eqno(6)$$
where the ``attempt frequency'' $\omega_0$ is the vibration frequency in the initial well
(here molecule bound to the solid surface) along the reaction coordinate (here desorption).
The probability for jumping over the barrier during the time period $t_0$ is
$P\approx wt_0$. If we take $t_0$ to be the time since the formation
of our solar system, $t_0 \approx 4.6\times 10^9 \ {\rm years}$ or 
$\approx 10^{18} \ {\rm s}$ and $\omega_0 \approx 10^{14} \ {\rm s}^{-1}$
from $P\approx 1$ we get ${\rm exp}(-\epsilon/ k_{\rm B}T ) \approx 10^{-32}$. 
If we assume the asteroid temperature
$T=200\ {\rm K}$ we get $\epsilon \approx 1.25 \ {\rm eV}$. 
We conclude that all loosely bound, and not so loosely bound,
atoms and molecules will all have desorbed from the surfaces of the asteroid particles. 
This includes all physisorbed molecules molecules such as methane or ethane (binding energies
$\sim 0.1-0.2 \ {\rm eV}$), molecules bound by hydrogen bonds (binding energies $\sim 0.1-0.4 \ {\rm eV}$),
most water molecules bound to glassy silica surfaces 
(typical binding energies $0.6-1.3 \ {\rm eV}$)\cite{water}, and even some chemisorbed 
molecules such as CO or NO on many metal and oxide surfaces (binding energies $\sim 1 \ {\rm eV}$).
[The bound water or hydroxyl-groups in certain minerals (e.g. phyllosilicates) are incorrectly
referred to as ``water''. They have nothing to do with ``free'' water that could form liquid bridges.]
This is likely to be the case even in the internal (cavity) regions of an asteroid since the
pores are likely to be connected to the vacuum region outside the asteroid via
open channels resulting from the asteroid porosity. In addition, most asteroids
have undergone several strong rearrangements of its fragments (break-up and reform) 
since their original formation, and hence it is likely
that all the surfaces of the solid particles have at least once been exposed to vacuum. 
[Calculations shows that asteroids spin-up to the point of break-up at a rate of at least 1 time in every
$\sim 10^6 \ {\rm years}$ (see Ref. \cite{not}).]
Hence all the solid particles can be expected
to have a very ``clean'' surface, and we expect no mobile molecules which 
could rearrange and form capillary bridges between
the particles. 

Here we note that it is possible that  free  water (or rather ice) exists in an
asteroid, possibly only in its interior, considering outer main-belt objects beyond the ``snow line'' 
and the blurry division between asteroids and comets. This water can form capillary ice bridges 
for temperatures well below the ice melting
temperature. This occurs mainly via sublimation and recondensation of water 
molecules in capillary bridges\cite{Gorb1,gundlach2018sintering}.

(d) Formation of chemical bonds.
Recent experiments and simulations
have found that nanoscale silica contacts exhibit aging due to the
formation of interfacial chemical bonds. First-principles calculations showed that a covalent
interfacial siloxane (Si-O-Si) bridge can
from in a thermally activated process from two silanol (Si-OH) groups on the opposing surfaces:
$${\rm Si-OH \ \ + \ \ Si-OH  \ \ \rightarrow  \ \ Si-O-Si \ \ + \ \ H_2O}$$
The activation energy involved for this reaction is about $1 \ {\rm eV}$ (see Ref. \cite{PRL1,PRL2,PRL3}).
and the reaction may thus occur even at $T=200 \ {\rm K}$ on the time scale of billion of years. Note also that
the force to break just one single chemical bond, $\sim \epsilon/a_0$, where $\epsilon \approx 1 \ {\rm eV}$
is the bond energy and $a_0 \approx 0.1 \ {\rm nm}$ the bond distance, is of order $1 \ {\rm nN}$
and hence similar to what we calculate for the van der Waals interaction (but less than for capillary
interaction) (see below).

Here we consider the adhesive interaction between particles with smooth surfaces.
We assume that the solids have large elastic modulus so that the elastic deformations of the solids
can be neglected. This problem was studied by Bradley\cite{bradley}
and Derjaguin\cite{Derjag}. (If elastic deformations cannot be neglected one must use
other theories such as the Derjaguin-Muller-Toporov DMT\cite{DMT} theory,
or the Johnson-Kendall-Roberts JKR\cite{jkr} theory, which hold for elastically stiff and 
soft solids, respectively.) For two spherical particles (radius $R_1$ and $R_2$)
in adhesive contact  theory  predicts the pull-off force\cite{Is}
$$F_{\rm ad} = 2 \pi \Delta \gamma R_{\rm eff}, \eqno(7)$$
where the effective radius $R_{\rm eff}$ is defined by
$${1\over R_{\rm eff}} = {1\over R_1}+{1\over R_2} .\eqno(8)$$
The work of adhesion $\Delta \gamma$ is the energy per unit surface area to separate 
two flat surfaces made from the same materials as in the sphere--sphere
contact problem. Fig. \ref{GluedAstro.1b.eps} illustrate the interaction between a sphere and a flat
in the rigid-solid limit.

Assuming the solids attract each other only via VDW interaction the work of adhesion
$$\Delta \gamma = {A\over 12 \pi r_0^2}, \eqno(9)$$
where $A$ is the Hamaker constant and $r_0$ is the separation between the flat surfaces 
(or between the spheres at the point of closest contact).
The (cut-off) distance $r_0$ is of atomic dimension. For amorphous ${\rm SiO_2}$ (silica) we will use
$A=6.5\times 10^{-20} \ {\rm J}$ and $r_0 = 0.3 \ {\rm nm}$ (see Ref. \cite{Alvo}). 
This gives $\Delta \gamma \approx 0.019 \ {\rm J/m^2}$
or $0.0012 \ {\rm eV/\AA^2}$. For a particle with the radius $R_1=10 \ {\rm \mu m}$ in contact with a flat 
($R_2=\infty$) this gives $F_{\rm ad}  \approx 1.2 \ {\rm \mu N}$.
However, $r_0 = 0.3 \ {\rm nm}$ may be too large. 
Israelachvili\cite{Is} suggest to use $r_0 \approx 0.165 \ {\rm nm}$ for many VDW systems. 
In Ref. \cite{Dunn} using $r_0 \approx 0.2  \ {\rm nm}$ was found to be consistent
with experiments for thermally oxidized polysilicone. For smooth surfaces the pull-off force
for any given $r_0$ can be obtained by scaling using that $F \sim r_0^{-2}$. Thus using $r_0 \approx 0.2 \ {\rm nm}$ 
as suggested in Ref. \cite{Dunn} would give a VDW pull-off force about $2$ times bigger than obtained above,
or about $F_{\rm ad} \approx 3 \ {\rm \mu N}$ for a particle with the radius $R_1=10 \ {\rm \mu m}$.

It is clear that $r_0$ is not a universal constant but depends on the lattice constants 
and crystal orientations of two molecular flat mineral crystals,
as well as the chemical peculiarities of the surface chemical groups.
To find more accurate values for common minerals in asteroids will be the subject of a future study. 
This problem is related to the problem of finding the reference plane for the VDW interaction for molecules above solid surfaces,
which was addressed in Ref. \cite{PZ1}.

In a humid atmosphere capillary bridges can form between the two spheres. If the spheres have smooth surfaces then
the pull-off force (neglecting the VDW interaction) is again given by (7) but with $\Delta \gamma = 2 \gamma$, 
where $\gamma$ is the fluid surface energy (or rather fluid-vapor interfacial energy). Here we have
assumed that the fluid wet the solids so the fluid-solid contact angle $\theta = 0$. For water\cite{new} 
$\gamma \approx 0.07 \ {\rm J/m^2}$
and $\Delta \gamma \approx 0.14  \ {\rm J/m^2}$ so for the case studied above the capillary pull-off force would be
$\sim 7$ times bigger then expected for the VDW interaction. We will show below that when the 
roughness of natural surfaces, e.g. produced by fracture, is included the difference in adhesion between 
VDW and capillary bridges becomes even larger, nearly a factor of $\sim 100$.

\begin{figure}[tbp]
\includegraphics[width=0.45\textwidth,angle=0]{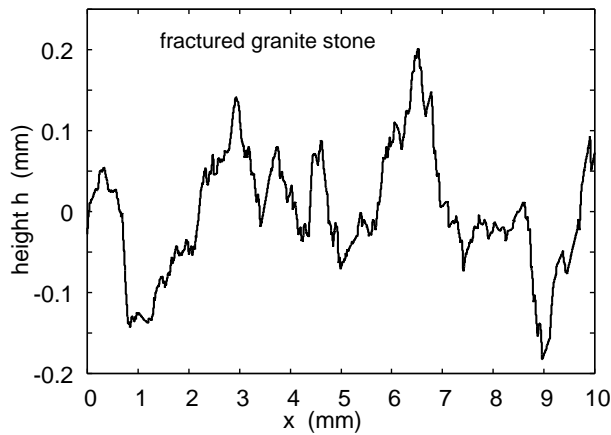}
\caption{
The surface height $h(x)$ along a $10 \ {\rm mm}$ strait line (coordinate $x$) on the fractured
granite surface.
}
\label{1x.2h.granite.eps}
\end{figure}

\begin{figure}[tbp]
\includegraphics[width=0.45\textwidth,angle=0]{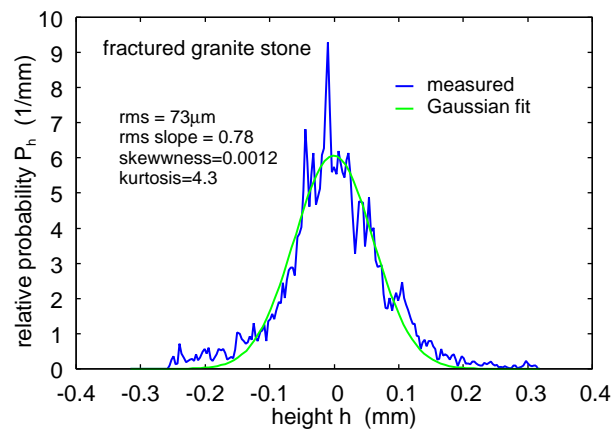}
\caption{
The probability distribution $P_h$ of surface heights.
}
\label{1h.2Ph.concrete.eps}
\end{figure}

\begin{figure}[tbp]
\includegraphics[width=0.45\textwidth,angle=0]{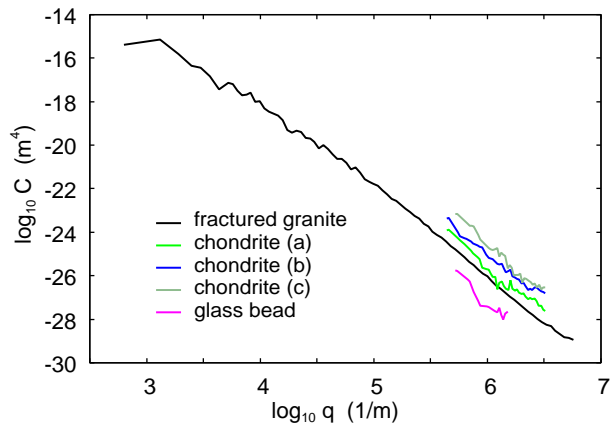}
\caption{
The surface roughness power spectrum as a function of the wave number (log-log scale).
For a fractured granite stone surface (black line) and for the meteorite particles (a)-(c)
(see Fig. \ref{TheParticles1.eps}) and for a glass bead. 
}
\label{1logq.2logC.all.astroid.NOTextrapolated.eps}
\end{figure}

\begin{figure}[tbp]
\includegraphics[width=0.65\textwidth,angle=0]{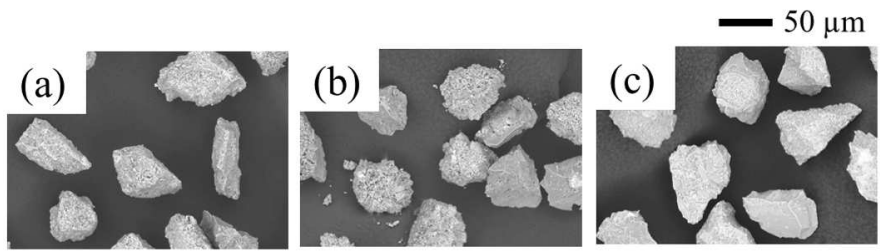}
\caption{
Three types of meteorite particles used in the adhesion study in Ref. \cite{Nag}. (a) and (b) are 
carbonaceous chondrite samples and (c) an ordinary chondrite particle. 
The meteorite particle power spectra shown in Fig. \ref{1logq.2logC.all.astroid.NOTextrapolated.eps} 
was obtained from the height profile $h(x,y)$ on square areas
of size $\sim 20 \ {\rm \mu m} \times 20 \ {\rm \mu m}$ on top of the particles. Adapted from \cite{Nag}.
}
\label{TheParticles1.eps}
\end{figure}

\vskip 0.3cm
{\bf 4 Surface roughness power spectra}

All surfaces of solids have surface roughness, and surfaces produced by fracture, as may be the case for
asteroid particles (or fragments) due to collisions between asteroids or due to the impact of meteorites, have usually 
large roughness which exhibit self-affine fractal behavior. This implies that if a surface area is magnified
new (shorter wavelength) roughness is observed which appear very similar to the roughness observed at smaller
magnification, assuming the vertical coordinate is scaled with an appropriate factor. 
The roughness profile $z=h({\bf x})$ of a surface can be written as a sum of plane waves ${\rm exp}(i{\bf q}\cdot {\bf x})$
with different wave vectors ${\bf q}$.
The wavenumber $q=|{\bf q}| = 2 \pi /\lambda$, where $\lambda$ is the wavelength of one roughness component.
A self affine fractal surface
has a two-dimensional (2D) power spectrum $C(q)\sim q^{-2(1+H)}$ (where $H$ is the Hurst exponent related to the 
fractal dimension $D_{\rm f} = 3-H$), which is a 
is a strait line with the slope $-2(1+H)$ when plotted on a log-log scale.
Most solids have surface roughness with the Hurst exponent $0.7 < H <1$ (see Ref. \cite{fractal}).
 
The most important information about the surface topography of a rough surface is the surface roughness power spectrum.
For a one-dimensional (1D) line scan $z=h(x)$ the power spectrum is given by
$$C_{\rm 1D} (q) = {1\over 2 \pi} \int_{-\infty}^\infty dx \ \langle h(x) h(0) \rangle e^{i q x}\eqno(10)$$
where $\langle .. \rangle$ stands for ensemble averaging.
For surfaces with isotropic roughness the 2D power spectrum $C(q)$ can be obtained directly 
from $C_{\rm 1D} (q)$ as described elsewhere \cite{Nayak,Carbone}.

Contact mechanics theory\cite{Johnson} shows that the contact between two solids 
with different surface roughness $h_1({\bf x})$ and $h_2({\bf x})$,
and different elastic properties (Young's modulus $E_1$ and $E_2$, and Poisson ratio $\nu_1$ and $\nu_2$)
can be mapped on a problem of the contact between an elastic half space 
(with the effective modulus $E$ and Poisson ratio $\nu$) with a flat surface, 
and a rigid solid with the combined surface roughness
$h({\bf x}) = h_1({\bf x})+h_2({\bf x})$. If the surface roughness on the two surfaces are uncorrelated 
then the surface roughness power spectrum of the rigid surface
$$C(q)=C_1(q)+C_2(q),\eqno(11)$$
where $C_1(q)$ and $C_2(q)$ are the power spectra of the original surfaces. The effective modulus of the
elastic solid is determined by
$${1\over E_{\rm eff}} = {1-\nu_1^2\over E_1}+{1-\nu_2^2\over E_2}\eqno(12)$$

For randomly rough surfaces, all the (ensemble averaged) information about the surface is contained in the 
power spectrum $C(q)$. For this reason
the only information about the surface roughness which enter in contact mechanics
theories (with or without adhesion) is the function $C(q)$.
Thus, the  (ensemble averaged) area of real contact, the interfacial stress distribution and the
distribution of interfacial separations, are all determined by $C(q)$\cite{Persson2,Prodanov,Carbone1}.

Note that moments of the power spectrum determines standard
quantities which are output of most stylus instruments and often quoted.
Thus, for example, the mean-square roughness amplitude $\langle h^2 \rangle$
and the mean-square slope $\langle (dh/dx)^2\rangle$ are 
is given by
$$\langle h^2 \rangle = 2 \int_0^\infty dq \ C_{\rm 1D}(q).$$
and
$$\left \langle \left ({dh\over dx} \right )^2 \right \rangle = 2 \int_0^\infty dq \ q^2 C_{\rm 1D}(q).$$
For isotropic roughness the 2D mean-square roughness amplitude is the same as the 1D mean-square roughness amplitude, but the
mean-square slope is a factor of 2 larger in the 2D case.

Using an engineering stylus instrument we have  
measured the roughness profile $z=h(x)$ of a granite surface produced by cracking a granite stone. 
The topography measurements was performed using Mitutoyo Portable Surface Roughness Measurement Surftest SJ-410 with a
diamond tip with the radius of curvature $R=1 \ {\rm \mu m}$, and with the tip-substrate repulsive force $F_{\rm N} = 0.75 \ {\rm mN}$.
The scan length $L=10 \ {\rm mm}$ and the tip speed $v=50 \ {\rm \mu m/s}$.
Fig. \ref{1x.2h.granite.eps} shows a $10 \ {\rm mm}$ long line scan with the rms-roughness roughness amplitude 
$h_{\rm rms} = 73 \ {\rm \mu m}$ and the (2D) rms slope $h'=0.78$. The probability distribution of surface heights is shown in Fig.
\ref{1x.2h.granite.eps} (blue line) which is rather well fitted by a Gaussian (green line), as expected for randomly rough surfaces.

Fig. \ref{1logq.2logC.all.astroid.NOTextrapolated.eps} shows the 2D surface roughness power spectrum of the granite surface, 
and of 3 meteorite particle surfaces (see Fig. \ref{TheParticles1.eps}) and of a glass bead surface.
The surface topography of the latter 4 surfaces was measured using an optical instrument\cite{Nag}, over small rectangular
surface areas on-top of the particles (which had diameters of order $50 \ {\rm \mu m}$). Optical methods often describe well
only the long wavelength roughness\cite{Boris}, and in the figure we only include the (long wavelength) region we trust.

\begin{figure}[tbp]
\includegraphics[width=0.45\textwidth,angle=0]{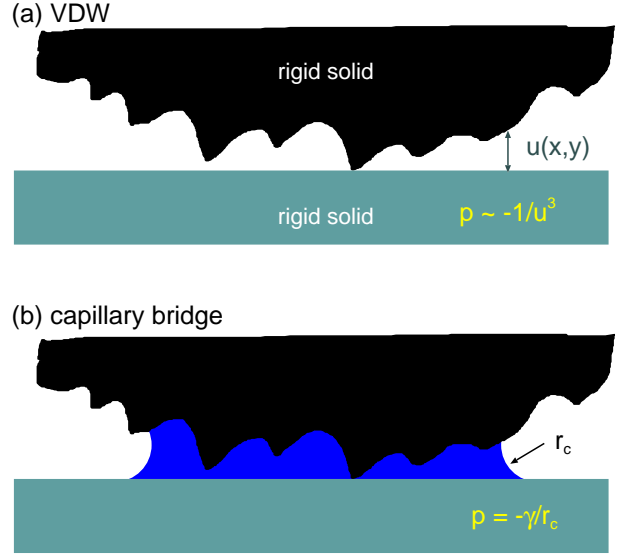}
\caption{
Two adhesion models. In (a) a particle with a rough
surface bind to a smooth surface with the van der Waals attraction.
In (b) the contact occurs in a humid atmosphere and a capillary bridge bind
the solids together. The capillary bridge is in thermal (kinetic) equilibrium with the 
surrounding gas of water molecules.
In both cases the solid objects are assumed perfectly rigid and they make contact in a single point.
}
\label{HowTo.eps}
\end{figure}

\begin{figure}[tbp]
\includegraphics[width=0.55\textwidth,angle=0]{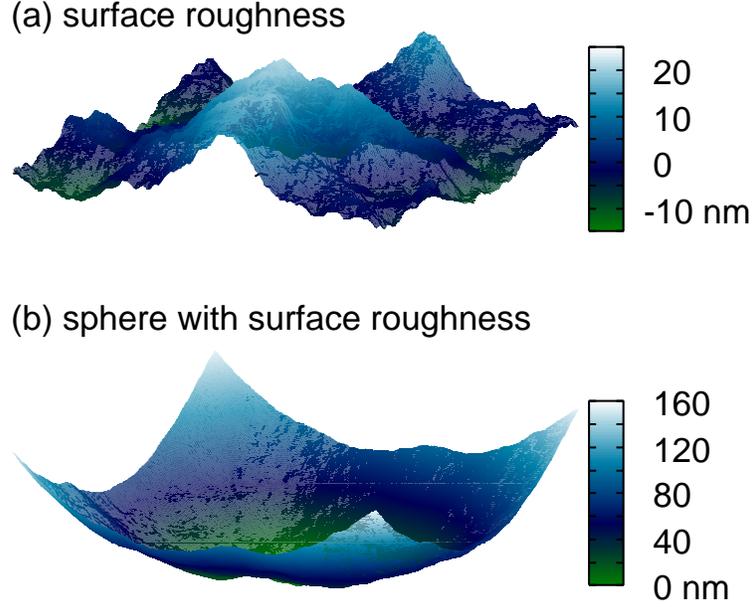}
\caption{
(a) Surface roughness generated using random number [see (14)]. 
(b) The surface roughness in (a) added on a nominally
spherical particle with the radius $R=632 \ {\rm nm}$. For the case when the roughness
on the granite surface is scaled by a factor of $s=0.25$.
}
\label{topography.X4.granite.s=0.25.ps}
\end{figure}

\vskip 0.3cm
{\bf 5 Adhesion of particles with surface roughness}

We consider the two different limiting adhesion models illustrated in Fig. \ref{HowTo.eps}.
In (a) a particle with a rough surface bind to a smooth surface with the van der Waals attraction.
In (b) the contact occur in a humid atmosphere and a capillary bridge bind
the solids together. The capillary bridge is in thermal (kinetic) equilibrium with the 
surrounding gas of water molecules determined by the relative humidity.
In both cases the solid objects are assumed perfectly rigid and they make contact in a single point.
In asteroids only the VDW interaction will exist because all weakly bound molecules, which could have formed
capillary bridges, will be desorbed (see above). However, when experiments are performed on the earth
there will usually be a non-zero humidity and capillary bridges will form at hydrophilic interfaces, e.g. between
most stone particles, or for silica glass in contact with silica glass, and this is the  
main reason why  we treat also liquid (water) bridges in this work. 

For the case of the VDW interaction we will assume that the solids attract each other with a stress
(or negative pressure) which depends on the local surface separation $u(x,y)$ as
$$p= - {A\over 6 \pi} {1\over u^3(x,y)}\eqno(13)$$
This equation is strictly valid only when $u(x,y)$ is a constant, 
and so small that retardation effects can be neglected, but we can use it approximately also when
the interfacial separation varies with the lateral coordinate $(x,y)$. The Hamaker constant $A$ can be calculated
from the dielectric properties of the solids using the Lifshitz theory of the van der Waals interaction\cite{Is,Parse}.
For silicon oxide (glass) is is approximately $A\approx 6.5 \times 10^{-20} \ {\rm J}$ and we will use this value for $A$
in the calculations below\cite{Is,good}. The minimum VDW separation, corresponding to atomic contact between the two solids,
is of order an atomic distance, and we will use $r_0=0.3 \ {\rm nm}$ in what follows\cite{use} (see the discussion above).

For capillary adhesion we assume that the fluid (water) wets the solid surfaces. Following Ref. \cite{cap5}, we put water at the interface
in all surface regions where the surface separation $u(x,y)$ is below the critical separation $2t+h_{\rm c}$, where
$t$ is the equilibrium thickness of the water film on the solid walls, and where $h_{\rm c}$ depends on the
humidity and is given by the Kelvin equation\cite{Kim1,Kim2,cap3,cap4,Bonn}. 

The (macroscopic) Kelvin equation relates the equilibrium curvature of the liquid-vapor interface with the vapor pressure, 
as derived by equating the chemical potentials between two bulk phases:
$${1\over r_{\rm eff}} = {k_{\rm B} T \over v \gamma} {\rm ln} {P_{\rm S}\over P}$$
where $r_{\rm eff}$ is the mean radius of curvature such that $1/r_{\rm eff} =1/r_1+1/r_2$ 
(where $r_1$ and $r_2$ are the two surface principal radius of curvatures) 
for the liquid meniscus. Here $k_{\rm B}$ is the Boltzmann constant, $T$ is the temperature,
$\gamma$ the surface tension of water,
$v=V/N$ the volume of a water molecule in water and $P/P_{\rm S}$ the relative humidity
($P_{\rm S}$ and $P$ are the saturated and actual water vapor pressure, respectively).

Both $2t$ and $h_{\rm c}$ depends on the humidity
and here we assume the relative humidity $\sim 40\%$. In this case
for water $h_{\rm c} \approx 2 \ {\rm nm}$ and (for amorphous silicone dioxide, silica) $2t \approx 2 \ {\rm nm}$ (see Ref. \cite{Kim1}). 
The (negative) pressure in the capillary bridges is given by the Laplace 
pressure $p\approx - \gamma /r_{\rm c}$, where $r_{\rm c}=h_{\rm c}/2$ is the radius of curvature of the capillary bridge
(see Fig. \ref{HowTo.eps}) at the vapor-fluid interface
(here we have neglected a small corrections denoted the Tolman length arising from the dependency of the surface tension
on the fluid curvature at the vapor-fluid interface)\cite{Is,Kim2}.

No two natural stone particle have the same surface roughness, and the adhesion force between two particles
will depend on the particles used. To take this into account we have generated particles 
(with linear size $L=2R$) with
different random surface roughness but with the same surface roughness power spectrum. 
That is, we use different realizations of the particle surface roughness
but with the same statistical properties. For each particle size we have generated
60 particles using different set of random numbers. The surface roughness
was generated as described in Ref. \cite{Rev} (appendix A) by adding plane waves with random phases 
$\phi_{\bf q}$ and with the amplitudes determined by the power spectrum:
$$h({\bf x}) = \sum_{\rm q} B_{\bf q} e^{i ({\bf q} \cdot {\bf x} + \phi_{\bf q})}\eqno(14)$$
where $B_{\bf q} = (2\pi /L) [C({\bf q})]^{1/2}$. We assume isotropic roughness so $B_{\bf q}$ and $C({\bf q})$ only depend on the
magnitude of the wavevector ${\bf q}$. Fig. \ref{topography.X4.granite.s=0.25.ps}(a) illustrate the surface roughness 
generated using (14), and in (b) we show a nominally spherical particle with the roughness given in (a).

\begin{figure}[tbp]
\includegraphics[width=0.45\textwidth,angle=0]{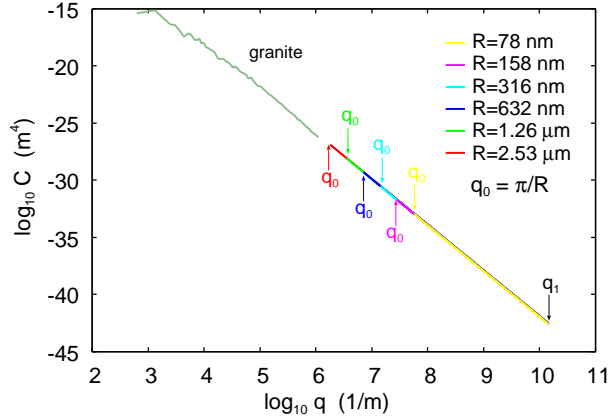}
\caption{
The curved line is measured power spectrum of the granite surface and the strait line
the (extrapolated) power spectra used for the (granite) particles with the radius indicated in the
figure. The small wavenumber (long wavelength) cut-off $q_0$ are indicated for each particle size
while the large wavenumber (short wavelength) cut-off $q_1$ is the same in all cases. The strait lines
has the slope $-4$ corresponding to the Hurst exponent $H=1$.
}
\label{1logq.2logC.used.concrete.manyRadius.1.eps}
\end{figure}

We have used nominally spherical particles with 6 different radius, where the radius increasing in steps of a factor of $2$ from
$R=78 \ {\rm nm}$ to $R=2.53 \ {\rm \mu m}$. The longest wavelength roughness which can occur on a particle with radius $R$
is $\lambda \approx 2R$ so when producing the roughness on a particle we only include the part of the power spectrum between
$q_0 < q < q_1$ where $q_0 = \pi /R$ and where $q_1$ is a short distance cut-off corresponding to atomic dimension
(we use $q_1 = 1.4\times 10^{10} \ {\rm m^{-1}}$). This is illustrated in Fig. \ref{1logq.2logC.used.concrete.manyRadius.1.eps} which shows the 
different short wavenumber cut-off $q_0$ used. We will refer to the particles with the power spectra shown in Fig.
\ref{1logq.2logC.used.concrete.manyRadius.1.eps} as granite particles because the power spectra used are linear extrapolation 
to larger wavenumber of the measured granite power spectrum. 

\begin{figure}[tbp]
\includegraphics[width=0.45\textwidth,angle=0]{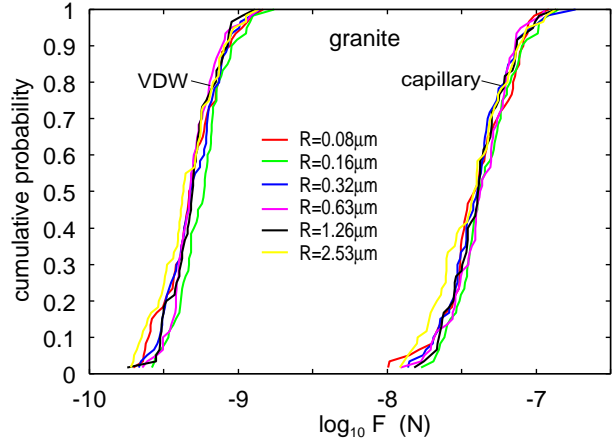}
\caption{
The cumulative probability for the pull-off force assuming capillary 
and VDW interaction. The probability distributions are obtained  
from 60 simulations for each particle radius. The 60 simulations use
60 different realizations of the particle surfaces topography but with the same power spectra.
The calculations are for the granite surface (scaling factor 1).
}
\label{1F.2PF.all.scale=1.eps}
\end{figure}

\begin{figure}[tbp]
\includegraphics[width=0.45\textwidth,angle=0]{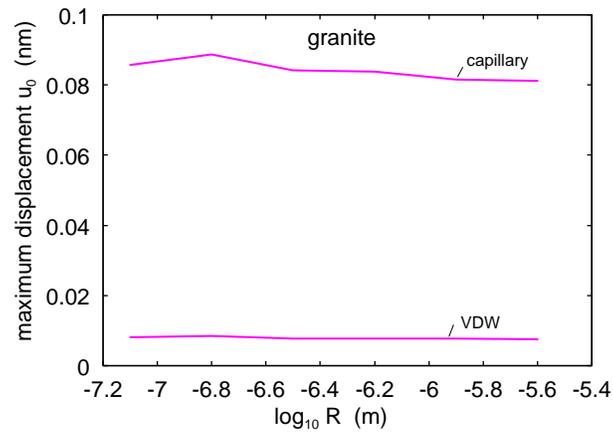}
\caption{
The maximum surface displacement (elastic deformation) in the contact region between
the particle and the substrate induced by the (adhesive) pressure distribution $p(x,y)$.
}
\label{1logR.2u.granite.c.and.VDW.eps}
\end{figure}

\begin{figure}[tbp]
\includegraphics[width=0.85\textwidth,angle=0]{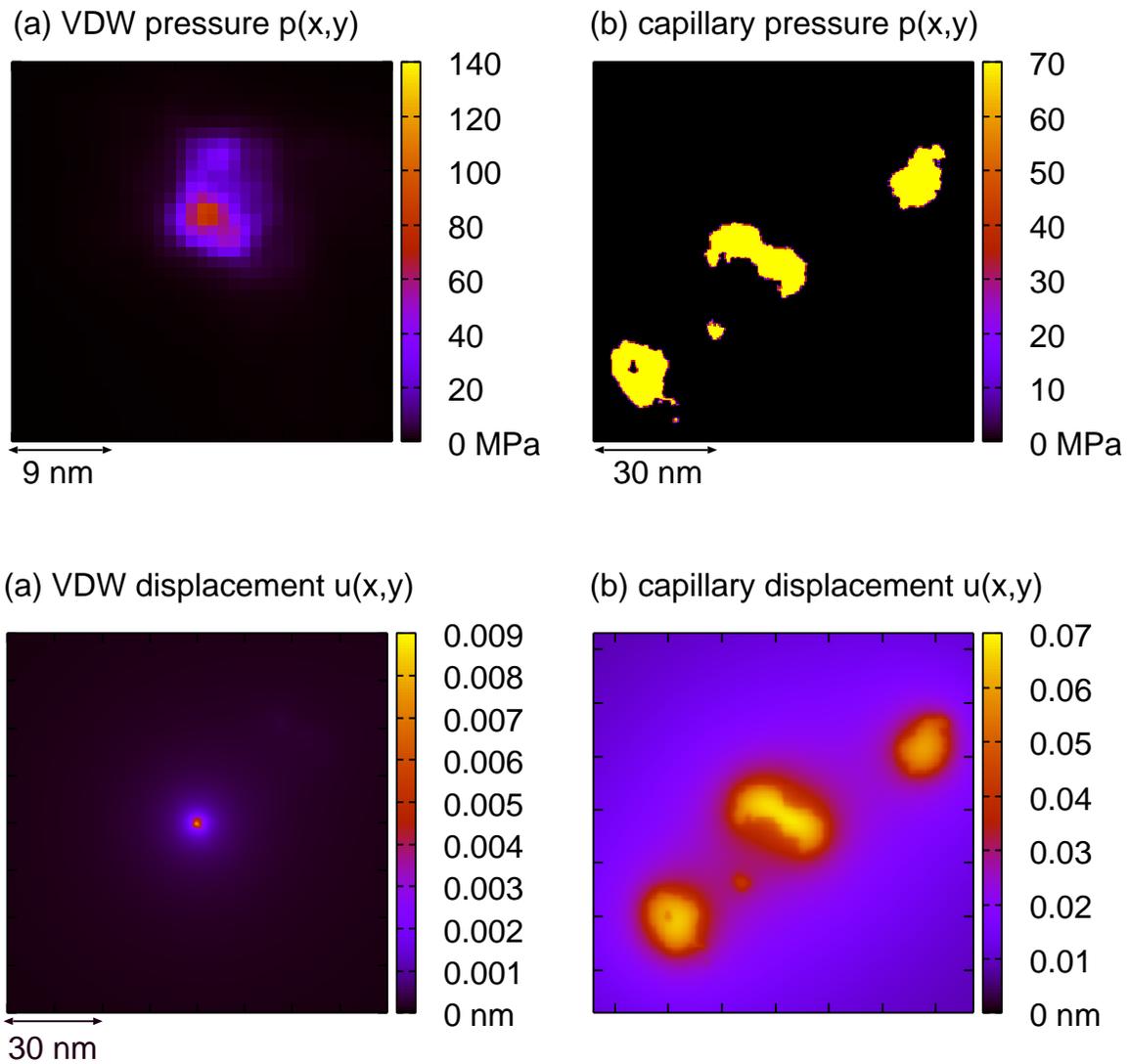}
\caption{
The contact pressure distribution and the surface displacement at the point of pull-off
assuming VDW interaction (a), (c) and capillary bridges (b), (d).
For a particle with radius $316 \ {\rm nm}$ and with the roughness power spectrum
of the granite surface.
}
\label{1xy.2u.and.p.for.c.VDW.flat.eps}
\end{figure}

\begin{figure}[tbp]
\includegraphics[width=0.45\textwidth,angle=0]{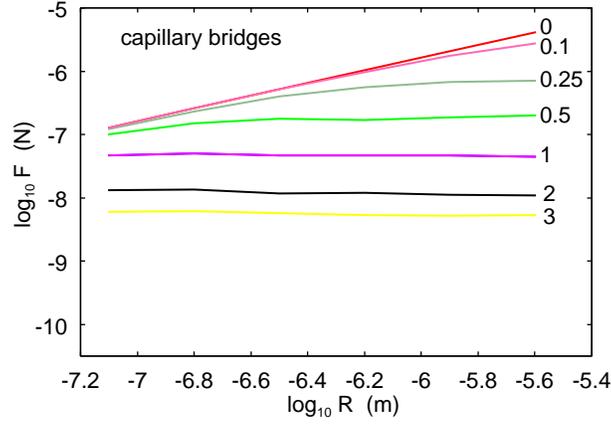}
\caption{
The calculated pull-off force as a function of the radius of the particle for several
surface roughness indicated by the scaling factor which change from 0 to 3.
The solids attract each other via capillary bridges (see text). 
}
\label{1logR.2logF.c.eps}
\end{figure}

\begin{figure}[tbp]
\includegraphics[width=0.45\textwidth,angle=0]{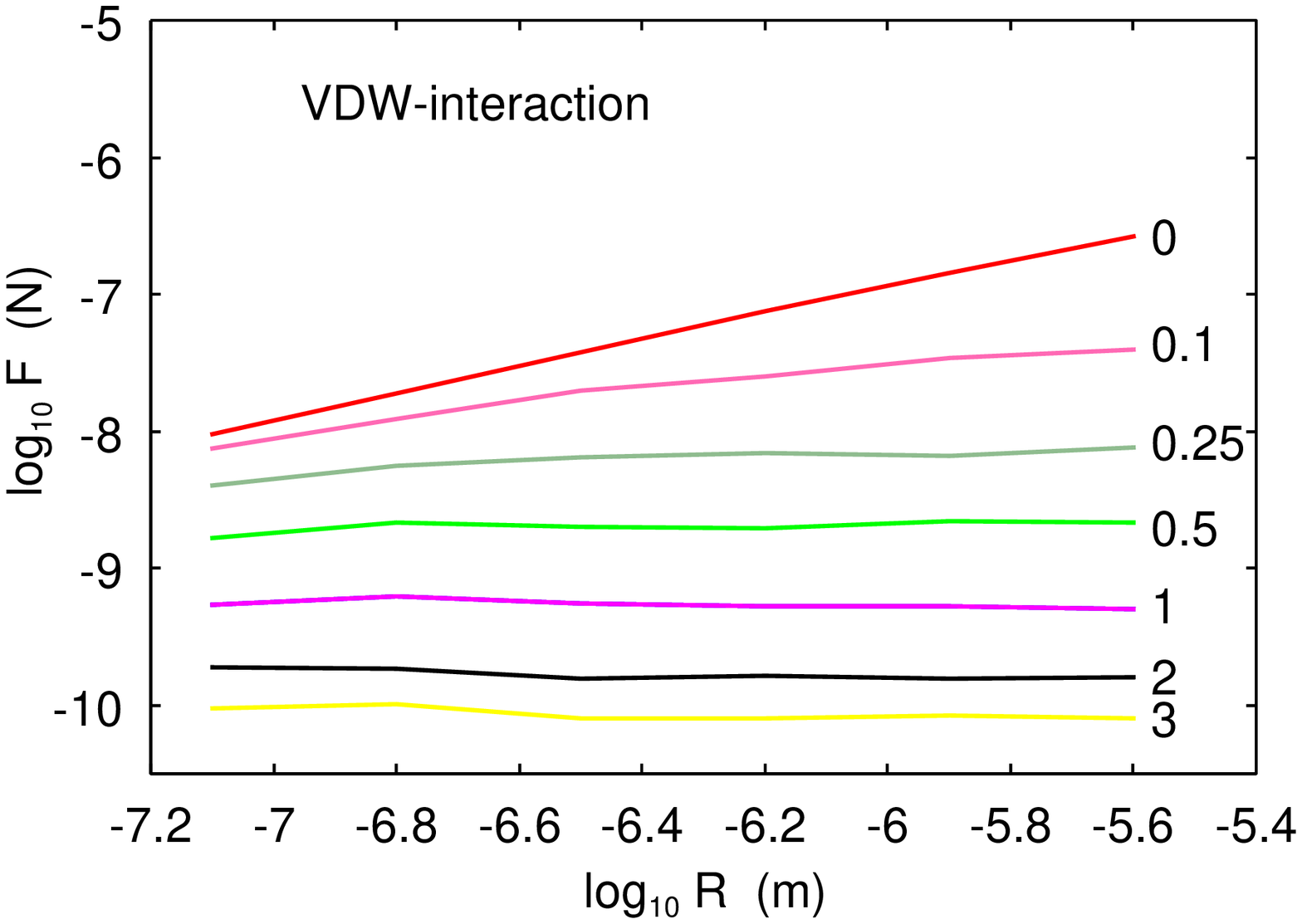}
\caption{
The calculated pull-off force as a function of the radius of the particle for several
surface roughness indicated by the scaling factor which change from 0 to 3.
The solids attract each other with the van der Waals interaction (see text).
}
\label{logR.logF.VDW.concrete.eps}
\end{figure}

\vskip 0.3cm
{\bf Numerical results for granite and meteorite particles}

We will now present numerical results for the adhesion of granite particles with the power spectra shown in Fig.
\ref{1logq.2logC.used.concrete.manyRadius.1.eps}. 
We will also consider particles with the same sizes as above but with 
larger and smaller surface roughness, obtained by scaling the
height $h(x,y)$ for the granite particles with scaling factors $s=0$ (smooth surface), $0.1$, $0.25$, $0.5$, $2$ and $3$.
Note that scaling $h(x,y)$ by a factor of $s$ will scale the power spectrum with a factor of $s^2$ but it will not change the
slope of the $C(q)$ relation on the log-log scale so the Hurst exponent (and the fractal dimension) is unchanged.

Fig. \ref{1F.2PF.all.scale=1.eps}
shows the cumulative probability for the pull-off force assuming capillary 
and VDW interactions. The probability distributions are obtained by using for each particle size
60 different surface roughness realizations with the same power spectra.
The calculations are for the granite surface (scaling factor $s=1$).
Note that the different curves corresponding to different particle radius 
gives nearly the same cumulative probability distribution  i.e.,
the pull-off force, and the  statistical fluctuations in the pull-off force, are 
nearly the same for all the particles. This imply that the pull-off force is independent of the
particle radius $R$, in sharp contrast to the linear dependency on $R$ for particles with smooth surfaces [see (7)].

Note that for a spherical granite particle with smooth surface
and the radius $R=2.5 \ {\rm \mu m}$ in contact with a flat granite surface the
pull-off force is [from (7)] $F=3.6\times 10^{-8} \ {\rm N}$ due to the VDW interaction, and 
$F=2.2\times 10^{-6} \ {\rm N}$ due to a capillarity.
However, for the real (rough) granite particles we get, after averaging over the 60 realizations of the surface roughness,
$F=5.1\times 10^{-10} \ {\rm N}$ and $F=4.4\times 10^{-8} \ {\rm N}$, respectively, i.e., 
smaller by factors of $71$ and $50$, respectively. For larger particles the
difference between smooth and rough surfaces is even bigger.

The adhesive pressure resulting from the VDW interaction and from capillary bridges will elastically
deform the surfaces. The deformation field $u(x,y)$ can be calculated from the theory of elasticity:
$$u({\bf x}) = {1\over \pi E_{\rm eff}}\int d^2x' {p({\bf x}') \over |{\bf x}-{\bf x}'|}$$
In the calculation of the surface displacement we have assumed that both solids are granite with the Young's elastic
modulus $E=64 \ {\rm GPa}$ and Poisson ratio $\nu =0.26$. Note that in this case the effective modulus $E_{\rm eff}$ is 
half of that of the granite [see (12)]\cite{Johnson}. 

Fig. \ref{1logR.2u.granite.c.and.VDW.eps} shows 
the maximum surface displacement (elastic deformation) in the contact region between
a particle and the substrate induced by the (adhesive) pressure distribution $p(x,y)$.
Note that the VDW interaction induces a deformation below $0.01 \ {\rm nm}$, which can be neglected
when calculating the pull-off force. The capillary bridge induce a larger deformation,
about $0.08 \ {\rm nm}$, but this displacement can also be neglected compared to the height of the capillary bridge,
$d_{\rm c} \approx 2 \ {\rm nm}$ in the present case.

Fig. \ref{1xy.2u.and.p.for.c.VDW.flat.eps} shows the contact pressure distribution $p(x,y)$ and the surface 
displacement $u(x,y)$ assuming VDW interaction (a), (c) and capillary bridges (b), (d).
The result is for a granite particle with radius $316 \ {\rm nm}$.
Note that for another nominally identical granite particle the pressure distribution, and the elastic
displacement, will look completely different because of the random nature of the surface roughness.
Note also that the contact pressure is always smaller than the macroscopic yield stress in compression, 
which is about $140 \ {\rm MPa}$ for granite. The microscopic yield stress, which is relevant here, might be even higher.

The displacement field $u(x,y)$ calculated above is for the case when there is a pull-off force acting on the particle which balance the
attractive force from the VDW or capillary interaction. In the equilibrium situation (no external force) the attractive force
is instead balanced by an equally strong repulsive force which in the present case would arise form a single contact point (or pixel)
(since we have assumed rigid solids). Of course, in reality the repulsive pressure will act over a finite region which cannot be smaller
than an atomic distance. In this region the pressure could be very high and one expect some local plastic flow or atom rearrangement.
Since this (plastically flattened) region would be very small in the present case the local deformations cannot be described using the 
(macroscopic) yield properties of the materials. Local plastic yielding may allow the solids to approach each other slightly
and may increase the pull-off force, but will not change the main conclusion of this study that, if the surface roughness
is big enough, the pull-off force is independent of the particle size. We note that in two recent studied plastic deformations was found to be
very important, but in these studies relative large normal force was used ($10 \ {\rm \mu N}$ preload in Ref. \cite{Pastew2} 
and $25-1857 \ {\rm mN}$ loading force in Ref. \cite{Bonn}), resulting in local plastic flow and enhanced pull-off forces.

Fig. \ref{1logR.2logF.c.eps} shows the capillary adhesion pull-off force as a function of the 
radius of the particle for several different
surface roughness amplitudes indicated by the scaling factors, which change from 0 to 3.
For the smooth surface (scaling factor $s=0$) the pull-off force increases linearly with
the radius of the particle, in agreement with the DMA theory prediction [see (7)].
However, for granite particles (scaling factor $s=1$), and for particles with even bigger
roughness ($s=2$ and $3$) as found for the meteorite particles (see Fig. \ref{1logq.2logC.all.astroid.NOTextrapolated.eps}), the
pull-off force is independent of the particle radius. 

Fig. \ref{logR.logF.VDW.concrete.eps} shows similar results as above for solids 
which attract each other with the van der Waals interaction.
For this case the dependency of the pull-off force on the scaling factor $s$ is similar as for capillary interaction,
but the pull-off force is much smaller.

\begin{figure}[tbp]
\includegraphics[width=0.45\textwidth,angle=0]{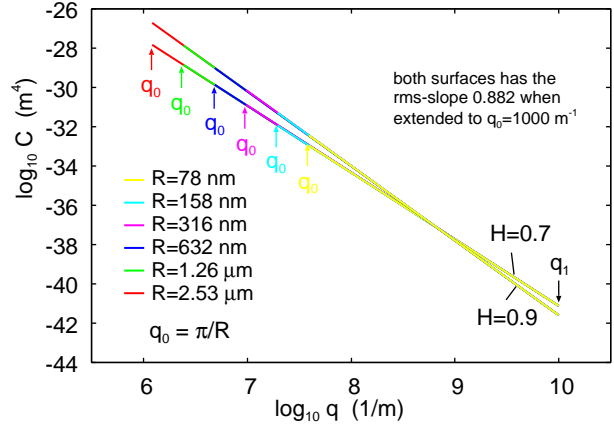}
\caption{
The power spectra of two set of particles with self-affine roughness with the Hurst exponent
$H=0.9$ and $H=0.7$ and with the radius indicated in the figure.
The small wavenumber (long wavelength) cut-off $q_0$ are indicated for each particle size
while the large wavenumber (short wavelength) cut-off $q_1$ is the same in all cases.
When the power spectra for the $H=0.9$ and $H=0.7$ cases are extrapolated to $q_0=1000 \ {\rm m^{-1}}$
the rms slope is $0.882$ for both surfaces while the rms-roughness amplitude differ. 
}
\label{1logq.2logC.H=0.9.H=0.7.slope.0.882.when.extended.to.q0=1000.eps}
\end{figure}

\begin{figure}[tbp]
\includegraphics[width=0.45\textwidth,angle=0]{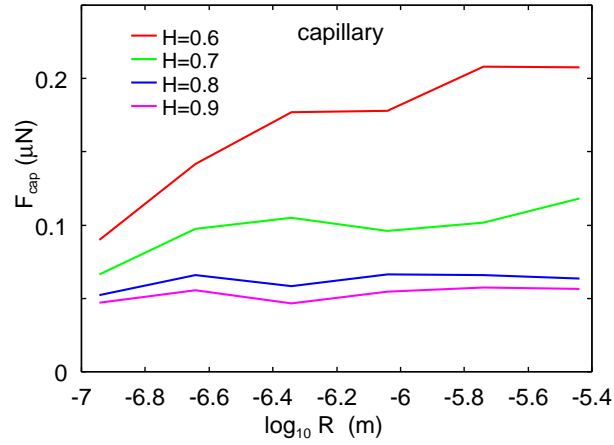}
\caption{
The calculated pull-off force as a function of the radius of the particle for several
Hurst exponents. 
For capillary bridges using
the surfaces with the power spectra shown in Fig. \ref{1logq.2logC.H=0.9.H=0.7.slope.0.882.when.extended.to.q0=1000.eps}
for $H=0.9$ and $H=0.7$ and similar power spectra for $H=0.8$ and $H=0.6$.
}
\label{1R.2F.c.all.H.eps}
\end{figure}

\begin{figure}[tbp]
\includegraphics[width=0.45\textwidth,angle=0]{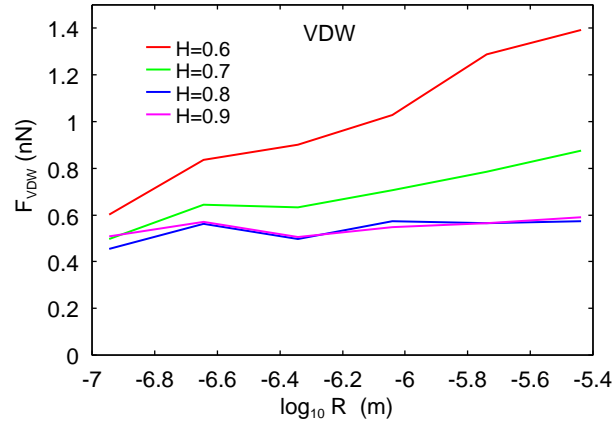}
\caption{
The calculated pull-off force as a function of the radius of the particle for several
Hurst exponents. 
For VDW interaction using
the surfaces with the power spectra shown in Fig. \ref{1logq.2logC.H=0.9.H=0.7.slope.0.882.when.extended.to.q0=1000.eps}
for $H=0.9$ and $H=0.7$ and similar power spectra for $H=0.8$ and $H=0.6$.
}
\label{1logR.2logF.VDW.eps}
\end{figure}

\begin{figure}[tbp]
\includegraphics[width=0.45\textwidth,angle=0]{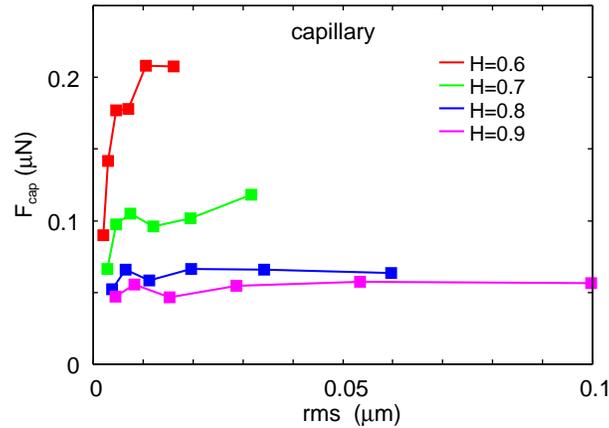}
\caption{
The calculated pull-off force as a function of the rms-roughness amplitude of the particle for several
Hurst exponents. In each case the particle radius increases from left to right: $R=0.114$, $0.228$, $0.455$, $0.910$,
$1.82$, $3.64 \ {\rm \mu m}$. 
For capillary bridges using
the surfaces with the power spectra shown in Fig. \ref{1logq.2logC.H=0.9.H=0.7.slope.0.882.when.extended.to.q0=1000.eps}
for $H=0.9$ and $H=0.7$ and similar power spectra for $H=0.8$ and $H=0.6$.
}
\label{1rms.2F.c.all.H.eps}
\end{figure}

\vskip 0.3cm
{\bf Numerical results for different Hurst exponents}

The numerical results presented above are for self-affine fractal surfaces with 
the Hurst exponent $H=1$, as found for granite. 
Here we will present results for other Hurst exponents.

We consider the adhesion for particles with self-affine fractal roughness
with the Hurst exponents $H=0.9$, $0.8$, $0.7$ and $0.6$.
In Fig. \ref{1logq.2logC.H=0.9.H=0.7.slope.0.882.when.extended.to.q0=1000.eps} 
we show the power
spectra of all the particles when $H=0.7$ and $H=0.9$. 
Self affine fractal surfaces have power spectra $C(q)$ which are strait lines on a log-log scale,
and the Hurst exponent determine the slope of the strait line [which equal $-2(1+H)$], 
but not the magnitude of the power spectra. For all Hurst exponents used here we have chosen
the magnitude of the power spectrum so that including all the roughness between $q_0=10^3 \ {\rm m^{-1}}$ and
$q_1=10^{10} \ {\rm m^{-1}}$ gives a surface with the rms-slope $0.882$. Note that the
rms-slope is mainly determined by the smallest wavelength roughness\cite{Rev} so for large  wavenumber
the magnitude of the power spectrum will be very similar in all cases.

In Fig. \ref{1R.2F.c.all.H.eps}
we show the pull-off force as a function of the radius of the particle for several
Hurst exponents assuming only capillary attraction. 
Fig. \ref{1logR.2logF.VDW.eps} shows similar results but
assuming only VDW interaction between the solid walls. Note that in both cases a pull-off force
is independent of the particle radius for $H=0.8$ and $H=0.9$. 
Studies have shown that almost all surfaces of practical interest have Hurst exponents between 
0.75 and 1 so for elastically stiff solids with large enough roughness, in almost all cases the pull-off force will be
independent of the particle radius. One exception to this may be surfaces with frozen capillary
waves\cite{Ref5,frozen1,frozen2} which have $H=0$, and may therefore exhibit a dependency of the pull-off force on the
particle radius.

The results presented above differ strongly from
the adhesion between elastically soft solids, where the adhesion force 
depends strongly on the long wavelength roughness,
which also has the biggest influence on the rms-roughness amplitude. 
For soft solids elastic deformations becomes very important,
and since most of the elastic deformation energy is stored in the deformation of the most long-wavelength roughness,
it is the longest wavelength roughness which often ``kill'' adhesion.
This has been shown theoretically\cite{adhe1,adhe2,Scaraggi,Mart1,Mart2,Mart3,Ciav}
and also observed in experiments\cite{Tiwari}. However, for elastically stiff solids, where the elastic deformations are
negligible, the situation is different. Thus when we increase the particle radius we also increase the rms-roughness
amplitude, but we have seen that the pull-off force in most cases does not depend on the particle radius. This is 
illustrated in Fig. \ref{1rms.2F.c.all.H.eps} for the case of capillary attraction. 
Note that the rms-roughness amplitude increases with a
factor of 23 (from $0.0044 \ {\rm \mu m}$ for the smallest particle to
$1.0 \ {\rm \mu m}$ for the largest particle) for the case when $H=0.9$, 
but the pull-off force does not change. We also note that there is no simple relation 
between the pull-off force and the surface rms-slope, as suggested 
recently\cite{Pastew,Pastew1,Pastew2}, since the rms-slope is nearly the same 
for all the studied cases with different 
Hurst exponents, while the pull-off force varies by more than a factor of $2$, 
and also exhibit different $R$-dependency (see Fig. \ref{1logR.2logF.VDW.eps}).

\begin{figure}[tbp]
\includegraphics[width=0.45\textwidth,angle=0]{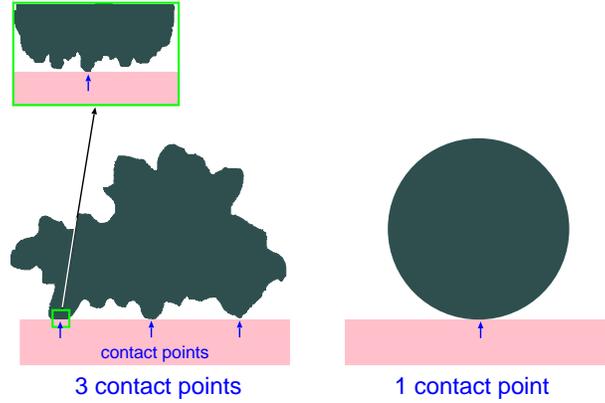}
\caption{A rigid particle in contact with a rigid substrate with a flat 
and perfectly smooth surface. 
An irregular shaped particle may make contact with the substrate in 
three points, separated by distances of order the diameter of the particle, 
while a perfectly smooth sphere may make contact in just
one point. A sphere with very small roughness may make contact with the substrate 
in three very closely spaced points.
}
\label{OneThree.eps}
\end{figure}

\begin{figure}[tbp]
\includegraphics[width=0.35\textwidth,angle=0]{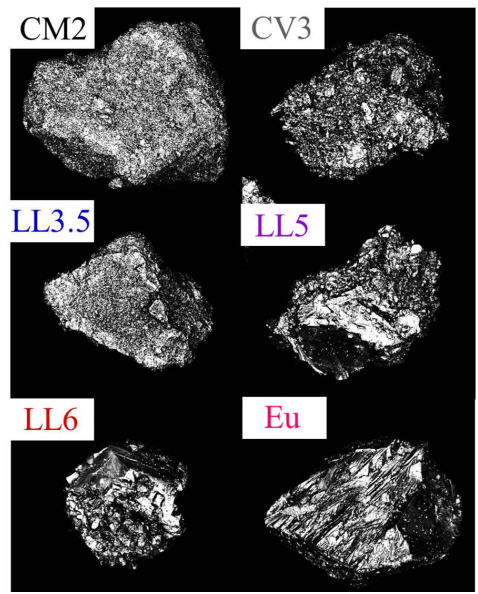}
\caption{
The meteorite particles used in the adhesion study in Ref. \cite{Nag}.
The particles CM2, CV3 and LL3.5 have the surface roughness power
spectra denoted by (a), (b) and (c) in Fig. \ref{1logq.2logC.all.astroid.NOTextrapolated.eps}.
Adapted from \cite{Nag}.
}
\label{3.eps}
\end{figure}

\begin{figure}[tbp]
\includegraphics[width=0.45\textwidth,angle=0]{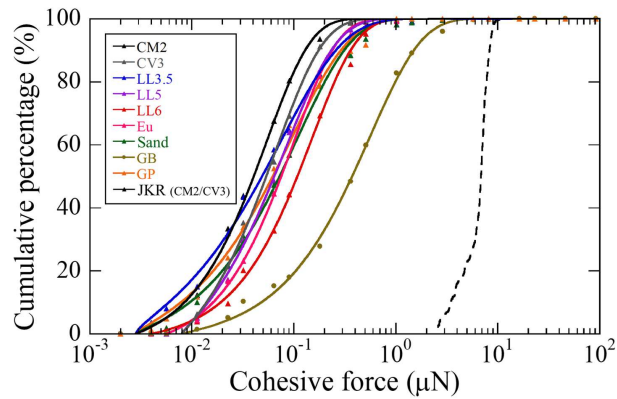}
\caption{
Cumulative percentage of the experimentally measured values of adhesive force.
The dashed curve is a model distribution of cohesive forces predicted by the
JKR theory in the case of silica sphere with the same size distribution as for the
meteorite particles. The curve denoted GB is for the glass beads. From Ref. \cite{Nag}.
}
\label{4.eps}
\end{figure}

\begin{figure}[tbp]
\includegraphics[width=0.25\textwidth,angle=0]{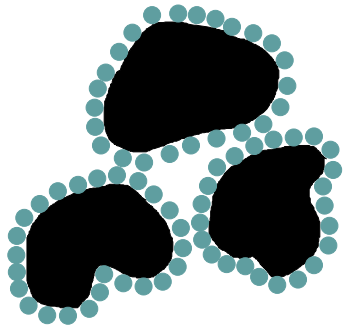}
\caption{
Small particles covering the surfaces of bigger particles will reduce
the effective adhesion between the big particles.
}
\label{TwoSizesDrug.eps}
\end{figure}

\begin{figure}[tbp]
\includegraphics[width=0.75\textwidth,angle=0]{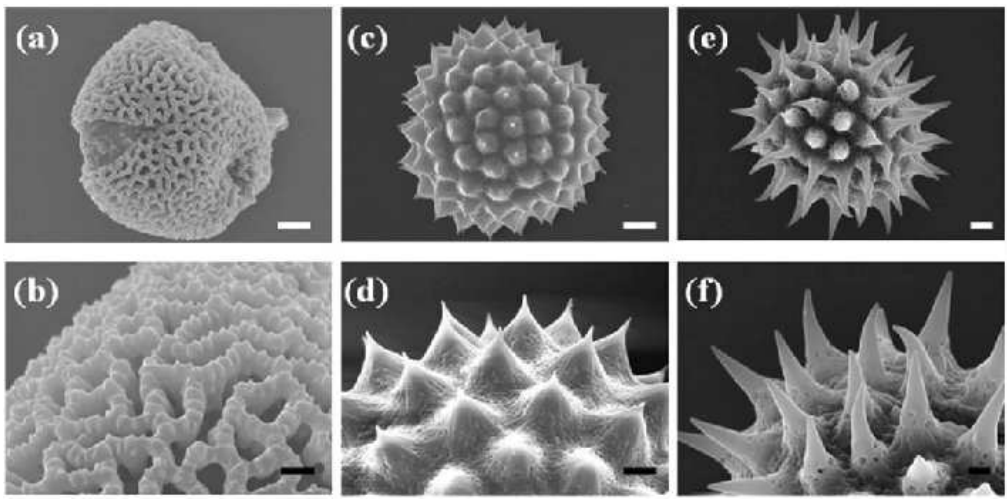}
\caption{
(a) Aggregates of desert dust particles. Adapted from \cite{desert}.
(b) SEM images of pollen particles: (a) and (b) are olive pollen; (c) and (d) are ragweed pollen; (e) and (f) are sunflower pollen. 
The white and black scale bar represent $2 \ {\rm \mu m}$ and $500 \ {\rm nm}$, respectively. From \cite{pollenpic} 
}
\label{DustPollen.eps}
\end{figure}

\begin{figure}[tbp]
\includegraphics[width=0.5\textwidth,angle=0]{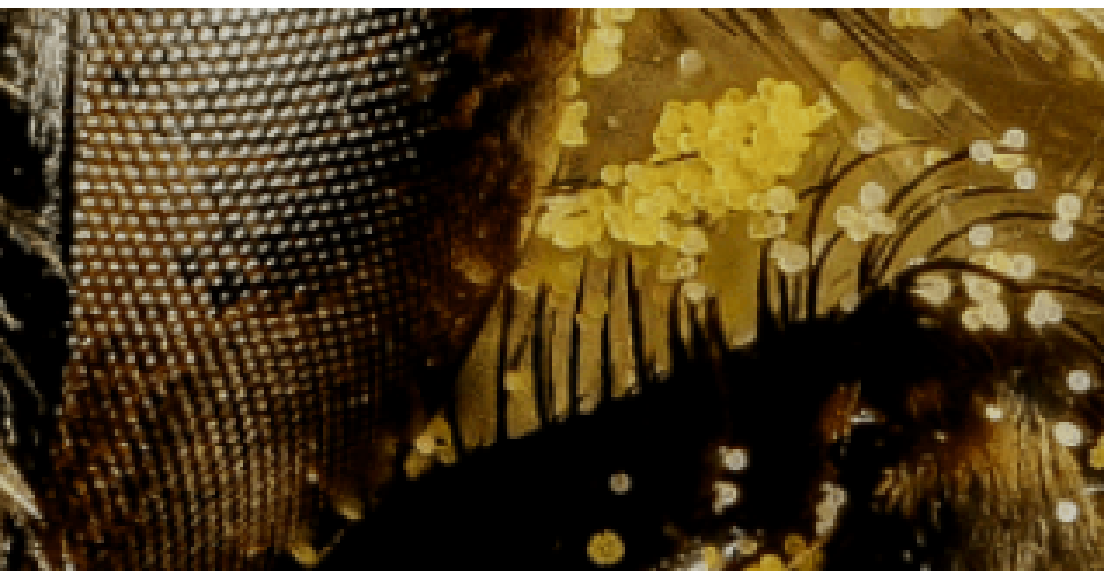}
\caption{
Optical picture of pollen attached to the hairs on a bee. Note that there is nearly no pollen attached to the
eye of the bee.
}
\label{pollenBee.eps}
\end{figure}

\vskip 0.3cm
{\bf 6 Discussion and comparison with experiments}

For the granite particles, the VDW pull-off force is $F\approx 0.6 \ {\rm nN}$ (see Fig. \ref{logR.logF.VDW.concrete.eps}).
For the particles with increased roughness, obtained with the scaling factors $s=2$ and $s=3$, the
pull-off force $F\approx 0.16 \ {\rm nN}$ and $\approx 0.09 \ {\rm nN}$, respectively. The actual pull-off
force of a particle from a flat surface may be higher because stable contact may require
three contact points, at least if the particles are big enough and exposed to a gravitational field
(see Fig. \ref{OneThree.eps}). In addition,
the VDW cut-off radius used above $r_0 = 0.3 \ {\rm nm}$ may be too large. Taking these factors into account approximately,
we obtain VDW pull-off forces for granite and the meteorite particles of order $0.3-2 \ {\rm nN}$. 

We have shown above that using $F\approx 1 \ {\rm nN}$ is consistent with measured rotation speeds 
of rubble pile asteroids. We note that if one instead would assume that $F$ is given by the standard
Derjaguin-Muller-Toporov (DMT) theory\cite{DMT} [which gives the pull-off force given by Eq. (7)], 
one would obtain particle pull-off forces which, even for the smallest particles of interest,
are $100-1000$ bigger then we calculate. But such large pull-off forces would be in strong disagreement
with the observed rotation speeds of asteroids.

In order for the small particles to act like a cement or glue for the bigger particles in asteroids it is important
that there are enough of them to fill out all cavity regions between the bigger particles. If this is not the case
then the effective adhesive force keeping the big fragments together will be strongly reduced. This effect is well known and
is used to reduce the adhesion between particles, e.g., in dry powder inhalers for drug delivery\cite{drug1}. 

Consider a solid made up by agglomeration of particles of equal size (diameter $d$).
The number of bonds between particles per unit volume scales as $1/d^3$ so decreasing the particle size
increases the cohesive strength of the solid. Consider now a mixture of small particles (the drug) and big
particles (the carrier particles). If there is just enough (or less) small particles to cover the big particles
(see Fig. \ref{TwoSizesDrug.eps}), and if the small particles bind strong enough to the big particles, then effectively 
the small particles only act to increase the roughness of the big particles. As a result the big particles
may touch each other only in a few locations where the the small particles attached to the big particles touch
each other. In this case, if the big particles would have relative smooth surfaces,
the coating with the small particles could strongly reduce the adhesion.
Thus while a (porous) solid block containing only the small particles
may exhibit strong cohesion, a solid containing big particles coated by the small particles may
exhibit negligible cohesion, and would easily flow when exposed to a small external force.

Note that increasing the adhesive forces between particles in an aggregate should raise the strength of the 
aggregate because each particle contact then requires more force for fracture. 
However, it is well known experimentally that strongly adhesive particles may lead to 
fluffy structures which contain fewer contacts and which are therefore weak, 
even though each individual particle contact may be stronger. Thus, adhesion can both increase and decrease 
the strength of aggregates, since the process of aggregation is inhibited by adhesion, whereas, 
the strength of the final aggregate is proportional to adhesion\cite{Kevin}. 

An important parameter influencing the packing density (1-porosity) of a granular medium is the Bond number defined
as the ratio between the adhesion force $F$ and the gravity force: $B=F/Mg$. Experiments have shown that if $B<<1$ 
a system of (of roughly equal sized) particles can pack well resulting in a low porosity solid, while if $B>>1$ the opposite is true
(see Ref. \cite{Capece}). If $F=1 \ {\rm nN}$ on the Earth where $g=9.81 \ {\rm m/s^2}$
we get $M\approx 10^{-10} \ {\rm kg}$ if $B=1$, corresponding to a particle with the diameter $D \approx 30 \ {\rm \mu m}$.
On a typical asteroid $g= 10^{-4}  \ {\rm m/s^2}$ giving $M\approx 10^{-5} \ {\rm kg}$ if $B=1$,
corresponding to a particle with the diameter $D \approx 1 \ {\rm mm}$. 

The pull-off forces measured for small particles on earth are usually larger than predicted above assuming
only the van der Waals interaction. This is often due to the formation of capillary bridges. Capillary
bridges form spontaneous in a humid atmosphere between contacting solids with a hydrophilic interface.
The influence of capillary bridges on the adhesion between small particles is well known from everyday
experience: dry sand may exhibit liquid-like flow, like in a sandglass (hourglass), while wet sand
particles can adhere, and as result one can building sand sculptures on the beach.

In a recent paper Nagaashi et al\cite{Nag} have measured, in ambient air (relative humidity $30-40\%$), 
the adhesion between meteorite particles 
(see Fig. \ref{3.eps}) (diameter $\sim 50 \ {\rm \mu m}$)
and a silica glass plate. They also studied the adhesion between silica glass beads and the same surface.
The adhesion force was measured by rotating the 
plate until the particles fly off. Due to fluctuations in the
particle surface roughness the particles fly off at different rotation speeds. Fig. \ref{4.eps}
shows the cumulative probability as a function of the pull-off force. The average (over many particles) 
pull-off force for the meteorite particles was $55$ (CM2), $68$ (CF3), $78$ (LL3.5), $87$ (LL5), 
$150$ (LL6), and $100 \ {\rm nN}$ (Eu) and for the glass bead $560  \ {\rm nN}$. We note that
the glass bead was relatively rough with the power spectrum shown below the granite power spectrum in
Fig. \ref{1logq.2logC.all.astroid.NOTextrapolated.eps}. The authors estimate from measured adsorption isotherms that approximately two water-vapor adsorption layers were present at the surface of the particles during the measurements of the cohesive forces, which is consistent with a capillary bridge between the contacting rough surfaces. 

For the granite particles we calculate the capillary pull-off force in a humid atmosphere
($\sim 40\%$ relative humidity) to be $F\approx 47 \ {\rm nN}$ (see Fig. \ref{1logR.2logF.c.eps}).
For the particles with increased roughness, obtained with the scaling factors $s=2$ and $s=3$, the
pull-off force $F\approx 13 \ {\rm nN}$ and $\approx 6 \ {\rm nN}$, respectively. The actual pull-off force may be 
nearly 3 times larger if three contact points occur, and if they break at the same time during pull-off,
which is possible because the bridges will elongate before breaking. Thus for the
meteorite particles we predict the pull-off force $20-150 \ {\rm nN}$, which is consistent with what is observed.
We note, however, that the Hamaker constant for the meteorite particles may be slightly different from 
granite, and the relative humidity may differ slightly (it was $30-40\%$ in the experiments of Ref. \cite{Nag}).

The power spectrum of the glass 
bead (see Fig. \ref{1logq.2logC.all.astroid.NOTextrapolated.eps}) is roughly one order of magnitude below
that of the granite corresponding to the scaling factor $s\approx 0.25$. For this case the pull-off force
depends on the radius of the particle but extrapolating the 
$s=0.25$ curve in Fig. \ref{1logR.2logF.c.eps} to $R\approx 25 \ {\rm \mu m}$ gives the pull-off force of about $700 \ {\rm nN}$.
Since the roughness of the glass bead is much smaller than for the meteorite particles it is possible that the
glass bead makes contact with the silica glass surface in just one point (see Fig. \ref{OneThree.eps}) in which case the
theory prediction would be in good agreement with the experimental observations.

There are many other applications of the results presented above. For example, the adhesion and removal of dust particles
from surfaces, e.g., wafer for electronic applications or from spacecraft\cite{shaking} , is an important topic. 
Thus Si wafer surfaces are usually contaminated with dust particles and impurities because of the various processes they go through. 
That's why wafer cleaning and surface conditioning are very important parts of wafer manufacturing. 
If the dust particles are elastically stiff, e.g., sand particles in dust storms (see Fig. \ref{TwoSizesDrug.eps}), 
the theory presented above is valid. 

Another interesting application where surface roughness may be of importance is pollen. Pollen surfaces are
highly structured, and often sharp structures point away from the surface (see Fig. \ref{TwoSizesDrug.eps}). These 
surface structures may be formed to reduce the adhesion between the pollen in order to avoid aggregation and the formation of compact blocks 
of pollen. However, in some cases a fluid (pollenkitt, an emulsion with water droplets in an oil) occur on the surface of the
pollen, at least in the valley between the sharp structures, but the biological reason for this fluid
is not fully understood\cite{Pollenkitt,Lin,Gorb2,Gorb3}. However, it is unlikely that the fluid occur to increase the adhesion 
(via capillary bridges) because this could be done more easily (by natural selection) with smoother surfaces. 

The pollen transported by bees have particular sharp and big surface structures. This may be in order to reduce
the adhesion to the bees body in particular to the relative smooth eyes.
Fig. \ref{pollenBee.eps}
shows an optical picture of pollen attached to the hairs on a bee. Note that there is nearly no pollen attached to the
eye of the bee. The eyes of the bees are covered by a high areal density of very thin hairs. This is likely to avoid 
adhesion of particles (not only pollen) to the bees eyes.

The elastic modulus of the pollen material is much smaller
than for stone particles, and in this case it may necessary 
to include the elastic deformations of the
pollen particles in order to obtain accurate interaction forces, 
and other contact mechanics properties.

\vskip 0.3cm
{\bf 7 Summary}

We have calculated the surface roughness power spectra for granite fragments, and of meteorite fragments, 
from the measured (stylus and optical) surface topography, and shown them comparable in the relevant 
parameters (Hurst exponent and scale). 
We have calculated the pull-off force due to van der Waals interaction, and due
to capillary bridges, between particles with self-affine fractal (random) roughness.
We have shown that the surface roughness, if big enough,
result in an interaction force which is independent of the size of the particles, in contrast to the
linear size dependency expected for particles with smooth surfaces. Thus, two stone fragments
produced by fracture, of linear size $100 \ {\rm nm}$, attract each other with the same non-gravitational 
force as two fragments of linear size $10 \ {\rm m}$. In this case the surface roughness reduces the pull-off 
force between micrometer sized particles by a factor of $\sim 100$, and even more for larger particles. 

This means that the dependence of cohesive strength of the granular medium on 
particle size is due to the increase in the number of particle-particle contacts (per unit area) alone.  
A decrease in particle size only increases the number of contacts without changing the strength 
of the particle-particle adhesive bond. 

Most asteroids with a diameter larger than $\sim 300 \ {\rm m}$ are
gravitational aggregates, i.e., consist of more than one solid object. All 
asteroids are rotating but almost all asteroids larger than $\sim 300 \ {\rm m}$
rotate with a period longer than $2.3 \ {\rm hours}$, which is the critical period 
at which a non-cohesive, self-gravitating aggregate will fail structurally. 
This indicates that there is nearly no adhesive interaction forces between the asteroid fragments.
We have show that this is due to the surface roughness of the asteroid particles.
However, observational data for asteroid rotation periods show that there are fast rotators, 
mostly of size $< 300 \ {\rm m}$, which rotate faster than the no-cohesion spin barrier predicts; 
this indicates that their integrity must 
be maintained by a small yield-strength of order $\sim 25 \ {\rm Pa}$.
A yield stress $\sigma_{\rm Y} \approx 25 \ {\rm Pa}$ can only be explained
if one assumes that there are enough small (of order a few micrometer) particles in asteroids 
to form a cement matrix (glue) between the bigger particles (fragments).  
We have shown that the pull-off force between stone fragments due to the van der Waals interaction  is of order
$\sim {\rm nN}$, which gives a cohesive yield stress $\sigma_{\rm Y}  \approx 25 \ {\rm Pa}$ 
if a matrix of micrometer sized particles surround the bigger stone fragments in asteroids;
this is consistent with observations.

In order for the small particles to act like a cement or glue for the bigger particles in asteroids it is important
that there are enough of them to fill out all cavity regions between the bigger particles. If this is not the case
then the effective adhesive force keeping the big fragments together will be strongly reduced. 

\vskip 0.5cm
{\bf Acknowledgments:}

We thank 
Y. Nagaashi, T. Aoki and A.M. Nakamura for the optical
images of the meteorite particles used in calculating the power spectra
shown in Fig. \ref{1logq.2logC.all.astroid.NOTextrapolated.eps}.

\end{document}